\documentclass[aps,onecolumn]{revtex4}
\usepackage{amsfonts}
\usepackage{amsmath}
\usepackage{amssymb,epsf}

\begin{document}

\title{Effects of energy dependent spacetime on geometrical thermodynamics
and heat engine of black holes: gravity's rainbow}
\author{B. Eslam Panah$^{1,2}$\footnote{%
email address: behzad.eslampanah@gmail.com}}
\affiliation{$^{1}$ Research Institute for Astronomy and Astrophysics of Maragha (RIAAM),
P.O. Box 55134-441, Maragha, Iran\\
$^2$ ICRANet, Piazza della Repubblica 10, I-65122 Pescara, Italy}

\begin{abstract}
Inspired by applications of gravity's rainbow in UV completion of general
relativity, we investigate charged topological black holes in gravity's
rainbow and show that depending on the values of different parameters, these
solutions may encounter with black hole solutions with two horizons, extreme
black hole (one horizon) or naked singularity (without horizon). First, we
obtain black hole solutions, calculate thermodynamical quantities of the
system and check the first law of thermodynamics. Then, we study the
thermodynamical behavior of the system including thermal stability and phase
transitions. In addition, we employ geometrical thermodynamics to probe
phase transition points and limits on having physical solutions. Finally, we
obtain heat engines corresponding to these black holes. The goal is to see
how black holes' parameters such as topological factor and rainbow functions
would affect efficiency of the heat engines.
\end{abstract}

\maketitle

\section{Introduction}

Since the introduction of general relativity, there has been an ongoing
attempt to modify gravity at the fundamental level. To name a few, one can
point out to Lovelock gravity \cite{LovelockI,LovelockII}, brane world
cosmology \cite{BraneI,BraneII}, scalar-tensor theories \cite%
{ScalarI,ScalarII}, $F(R)$ gravity \cite{F(R)I,F(R)II,F(R)III,F(R)IV},
massive gravity \cite{MassiveI,MassiveII,MassiveIII}, and also gravity's
rainbow \cite{Smolin,Magueijo}. Among these modifications of Einstein
gravity, gravity's rainbow seems to be a promising candidate in dealing with
Ultra-Violet divergences (UV).

To introduce gravity's rainbow, one should first understand the concepts of
doubly special relativity. In special relativity, an upper limit of the
velocity of light is imposed on particles' velocities. Following the same
principle, one can also consider an upper limit on energy of the particles
as well. This upper limit is the Plank energy. In the literature, this is
called doubly special relativity (DSR) \cite%
{Amelino,Amelino1,Amelino2,Amelino3}. Indeed, DSR is an extension of special
relativity where there are two fundamental upper bounds on properties of
particles: the velocity of light and Planck energy. The DSR is in the
non-curved space. If we generalize such theory to curved spacetime and
include the gravity, we will have doubly general relativity or gravity's
rainbow \cite{Smolin,Magueijo}. In gravity's rainbow, the test particles
with various energy experience the gravity differently. In other words,
gravity has an effective behavior on particles determined by their energy.
Going back to DSR, in order to incarnate the upper limit on energy
particles, one should use a nonlinear Lorentz transformation in momentum
space. This implies a deformed Lorentz symmetry such that the usual
dispersion relation in special relativity may be modified by Planck scale
correction. It should be noted that it was speculated that the Lorentz
infraction or deformation may be an essential property in constructing a
quantum theory of gravity.

The modified version of the energy-momentum dispersion is given by
\begin{equation}
E^{2}f^{2}\left( E/E_{p}\right) -p^{2}g^{2}\left( E/E_{p}\right) =m^{2}
\end{equation}%
where $m$ and $E$ are mass and energy of a test particle, respectively. Also
$E_{p}$ is the Planck energy scale. For simplicity, we rename the following
ratio $\varepsilon =E/E_{p}$. The value of this ratio is always less than
one, because the energy of a test particle can not be larger than the Planck
energy \cite{Peng}. The functions $f\left( \varepsilon \right) $ and $%
g\left( \varepsilon \right) $ are called rainbow functions and satisfy the
following equation
\begin{equation}
\lim_{\varepsilon \rightarrow 0}f\left( \varepsilon \right) =1\
,\lim_{\varepsilon \rightarrow 0}g\left( \varepsilon \right) =1,
\end{equation}%
these limits are originated from infrared (IR) limit where the usual general
relativity must be recovered. Studies conducted with consideration of
gravity's rainbow proven to be fruitful. Among the achievements of gravity's
rainbow, one can point to providing possible solutions for information
paradox \cite{AliFM,Gim}, absence of black hole production at LHC \cite%
{AliFK}, and the existence of remnants for black holes after evaporation
\cite{Ali}. Cosmologically speaking, consideration of gravity's rainbow
could remove the big bang singularity \cite%
{NonsigularI,NonsigularII,NonsigularIII}. Also the initial singularity
problem \cite{CosmologyI} and stability of Einstein static universe in
gravity's rainbow have been studied \cite{CosmologyII}. In the context of
astrophysics, it was shown that the maximum mass of neutron stars \cite%
{HendiBEP,EslamPanah} and white dwarfs \cite{White dwarfs} in the presence
of gravity's rainbow, are an increasing function of rainbow functions (see
Refs. \cite{GarattiniTOV,GarattiniTOVI}, for another properties of neutron
stars and white dwarfs). In the context of gravity, it was pointed out that
Horava-Lifshitz gravity can be related to the gravity's rainbow through
suitable scaling for the energy functions \cite{Garattini}. In the context
of black holes, several studies focusing on effects of gravity's rainbow on
thermodynamics of black holes in general relativity were done in Refs. \cite%
{BHRI,BHRII,BHRIII,BHRIV,BHRV,BHRVI,BHRVII}. The influence of gravity's
rainbow to the global Casimir effect around a static mini black hole at zero
and finite temperature has been studied in Ref. \cite{Alencar}.
Consideration of gravity's rainbow with modified gravity such as $F(R)$
theories of gravity \cite{GarattiniJCAP,HendiFR}, massive gravity \cite%
{MassiveHEP}, Gauss--Bonnet gravity \cite{HendiF,HendiBB}, Lovelock gravity
\cite{LovelockH}, and also, dilaton gravity \cite{HendiDilatonI} were also
done.

Among the different solutions in gravity, the black holes have been an
interesting subject that have attracted a lot of attention over past $100$
years old. The discovery of the gravitational waves on one hand and
thermodynamical properties of the black holes on the other hand \cite%
{Therm1,Therm2,Therm3}, have made black hole an interesting platform for
investigating the nature of gravity. In addition, the introduction of the
adS/CFT duality and string theory inspired applications of the black holes
provided more reasons to investigate the black holes \cite%
{AdS1,AdS2,AdS3,AdS4,AdS5,AdS6}. Considering these issues, we study black
hole solutions in gravity's rainbow here.

Among the recent thermodynamical advances in the field of black holes
thermodynamics, one can point out to extended phase space and geometrical
thermodynamics. In the extended phase space, one consider the cosmological
constant in adS black holes to be a thermodynamical quantity known as
pressure \cite{Kastor,kubuiznak}. Such consideration results into
introduction of a van der Waals behavior, the reentrant of phase transition
\cite{reentrant1,reentrant2}, existence of the triple point \cite%
{triple1,triple2} and possibility of having classical heat engine \cite%
{Johnson,JohnsonI,JohnsonII,Bhamidipati,ChargeHeat,ConformalHeat,Sadeghi,Mo,Setare,Hendiheat,Accelerating,HeatmoreI,HeatmoreII,HeatmoreIII,HeatmoreIV}%
. The geometrical thermodynamics approach uses the thermodynamical
quantities to built a metric which describes thermal phases of the
corresponding black hole \cite%
{WeinholdI,WeinholdII,RuppeinerI,RuppeinerII,QuevedoI,QuevedoII,HPEM}. The
thermodynamical behavior is understood by Ricci scalar of the metric.
Changes in Ricci scalar's sign and its divergencies are employed to depict
the thermal phases of black holes in this approach. The aim is to have
thermodynamical structure of the black holes described by Rimennian
calculus. So far, different methods are proposed for constructing the
thermodynamical metric which among them, one can point out to Weinhold \cite%
{WeinholdI,WeinholdII}, Ruppeiner \cite{RuppeinerI,RuppeinerII}, Quevedo
\cite{QuevedoI,QuevedoII} and HPEM \cite{HPEM}.

In this paper, we intent to investigate thermodynamical properties of the
black holes in the presence of gravity's rainbow. Here, we address the
modifications imposed on thermal stability conditions and possible bounds on
having physical solutions. In addition, we regard geometrical thermodynamic
approach to investigate phase transitions of the solutions. Next, we
construct heat engines inspired by these black holes and show how the
rainbow functions would modify the efficiency of these engines. The paper
will be concluded by some closing remarks.

\section{Black holes in gravity's rainbow}

Here, we start with the $4$-dimensional action of Einstein gravity in the
presence of cosmological constant which depends on the energy ($\Lambda
(\varepsilon )$) with an abelian $U(1)$ gauge field as
\begin{equation}
I=-\frac{1}{16\pi G(\varepsilon )}\int d^{4}x\sqrt{-g}\left( R-2\Lambda
(\varepsilon )-F\right) ,  \label{action}
\end{equation}%
where $R$ is the scalar curvature. $F=F_{\mu \nu }F^{\mu \nu }$ is the
Maxwell invariant, in which $F_{\mu \nu }=\partial _{\mu }A_{\nu }-\partial
_{\nu }A_{\mu }$ is the Faraday tensor, and also $A_{\mu }$ is the gauge
potential.

Variation of the action (\ref{action}) with respect to the metric ($g_{\mu
\nu }$) and the Faraday tensors ($F_{\mu \nu }$) leads to the following
field equations
\begin{eqnarray}
G_{\mu \nu }(\varepsilon )+\Lambda (\varepsilon )g_{\mu \nu } &=&8\pi
G(\varepsilon )T_{\mu \nu },  \label{Field equation} \\
&&  \notag \\
\nabla _{\mu }F^{\mu \nu } &=&0,  \label{Maxwell}
\end{eqnarray}%
where
\begin{equation}
T_{\mu \nu }=2F_{\mu \lambda }F_{\nu }^{\lambda }-\frac{1}{2}g_{\mu \nu }F,
\label{EMtensor}
\end{equation}%
$G_{\mu \nu }(\varepsilon )$ is Einstein's tensor which is assumed to be
energy dependent. Since we are working in natural units, we set $8\pi
G(\varepsilon )=1$.

We consider a topological $4$-dimensional static energy dependent spacetime
with the following form
\begin{equation}
ds^{2}=-\frac{\psi (r,\varepsilon )}{f^{2}\left( \varepsilon \right) }dt^{2}+%
\frac{1}{g^{2}\left( \varepsilon \right) }\left[ \frac{dr^{2}}{\psi
(r,\varepsilon )}+r^{2}d\Omega ^{2}\right] ,  \label{metric}
\end{equation}%
where $\psi (r,\varepsilon )$ is the metric function. Also, in the above
equation, $d\Omega ^{2}$ is given by
\begin{equation}
d\Omega ^{2}=\left\{
\begin{array}{cc}
d\theta ^{2}+\sin ^{2}\theta d\varphi ^{2} & k=1 \\
d\theta ^{2}+d\varphi ^{2} & k=0 \\
d\theta ^{2}+\sinh ^{2}\theta d\varphi ^{2} & k=-1%
\end{array}%
\right. ,
\end{equation}

It is notable that the constant $k$ indicates that the boundary of $%
t=constant$ and $r=constant$ can be elliptic ($k=1$), flat ($k=0$) or
hyperbolic ($k=-1$) curvature hypersurface.

In order to obtain electrically charged black holes in gravity's rainbow, we
consider a radial electric field which its related gauge potential is in the
following form
\begin{equation}
A_{\mu }=h(r,\varepsilon )\delta _{\mu }^{t}.
\end{equation}

Using the metric (\ref{metric}) with the Maxwell equations (\ref{Maxwell}),
we can find the following differential equation
\begin{equation}
rh^{\prime \prime }(r,\varepsilon )+2h^{\prime }(r,\varepsilon )=0,
\label{diffh(r)}
\end{equation}%
in which the prime and double prime, respectively, are the first and the
second derivatives with respect to $r$. One finds the solution of the
equation (\ref{diffh(r)}) as
\begin{equation}
h(r,\varepsilon )=-\frac{q(\varepsilon )}{r},
\end{equation}%
where $q(\varepsilon )$ is a parameter which is related to the electric
charge. It is worthwhile to mention that the electromagnetic field tensor is
$F_{tr}=\partial _{t}A_{r}-\partial _{r}A_{t}=\frac{q(\varepsilon )}{r^{2}}$%
, which depends on the energy-dependent electrical charge ($q(\varepsilon )$%
) and $r$.

Considering the introduced metric (\ref{metric}) and the field equations (%
\ref{Field equation}), we want to obtain exact solutions for the metric
function $\psi (r,\varepsilon )$. We obtain the following differential
equations
\begin{eqnarray}
eq_{tt} &=&eq_{rr}=g^{2}\left( \varepsilon \right) \left[ r^{3}\psi ^{\prime
}(r,\varepsilon )+r^{2}\left( \psi (r,\varepsilon )-k\right)
+q^{2}(\varepsilon )f^{2}(\varepsilon )\right] +\Lambda (\varepsilon )r^{4},
\label{eq1} \\
&&  \notag \\
eq_{\theta \theta } &=&eq_{\varphi \varphi }=g^{2}\left( \varepsilon \right)
\left[ 2r^{3}\psi ^{\prime }(r,\varepsilon )+r^{4}\psi ^{\prime \prime
}(r,\varepsilon )-2q^{2}(\varepsilon )f^{2}(\varepsilon )\right] +2\Lambda
(\varepsilon )r^{4},  \label{eq2}
\end{eqnarray}%
where $eq_{tt}$, $eq_{rr}$, $eq_{\theta \theta }$ and $eq_{\varphi \varphi }$
are components of $tt$, $rr$, $\theta \theta $ and $\varphi \varphi $ of
field equation (\ref{Field equation}), respectively. Considering eqs. (\ref%
{eq1}) and (\ref{eq2}), after some calculations, one can obtain the
following metric function
\begin{equation}
\psi \left( r,\varepsilon \right) =k-\frac{m_{0}(\varepsilon )}{r}-\frac{%
\Lambda (\varepsilon )r^{2}}{3g^{2}\left( \varepsilon \right) }+\frac{%
q^{2}(\varepsilon )f^{2}\left( \varepsilon \right) }{r^{2}},  \label{BH}
\end{equation}%
where $m_{0}(\varepsilon )$\ is integration constant related to the total
mass of the black hole.

\begin{figure}[tbp]
$%
\begin{array}{c}
\epsfxsize=10cm \epsffile{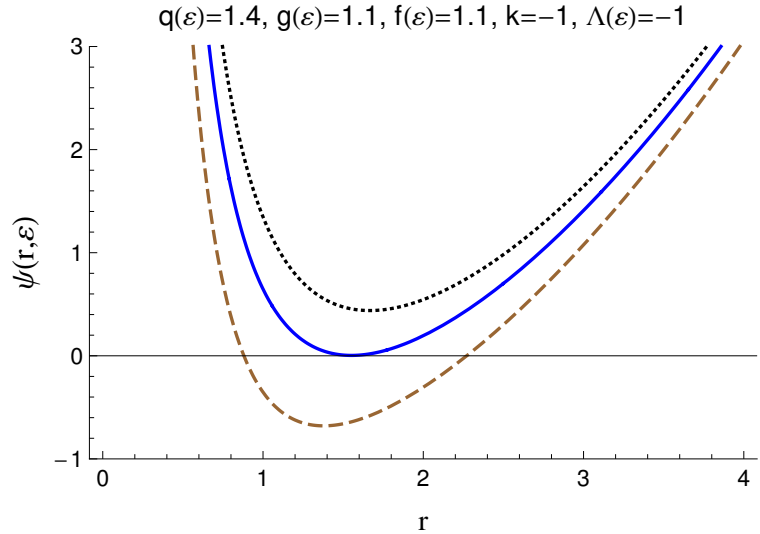}%
\end{array}
$%
\caption{$\protect\psi (r,\protect\varepsilon )$ versus $r$ for $m_{0}(%
\protect\varepsilon )=0.3$ (dotted line), $m_{0}(\protect\varepsilon )=1.0$
(continuous line) and $m_{0}(\protect\varepsilon )=2.0$ (dashed line).}
\label{Fig1}
\end{figure}

In order to investigate the geometrical structure of these solutions, we
first look for the essential singularity(ies). The Ricci and Kretschmann
scalars can be written as
\begin{eqnarray}
R &=&4\Lambda (\varepsilon )  \label{Ricci} \\
&&  \notag \\
R_{\mu \nu \lambda \kappa }R^{\mu \nu \lambda \kappa } &=&\frac{8\Lambda
^{2}(\varepsilon )}{3}+\frac{12m_{0}^{2}(\varepsilon )g^{4}(\varepsilon )}{%
r^{6}}-\frac{48m_{0}(\varepsilon )q^{2}(\varepsilon )f^{2}\left( \varepsilon
\right) g^{2}(\varepsilon )}{r^{7}}  \notag \\
&&+\frac{56q^{4}(\varepsilon )f^{4}\left( \varepsilon \right)
g^{4}(\varepsilon )}{r^{8}},  \label{Kretschmann}
\end{eqnarray}%
where our calculations confirm that, there is a curvature singularity at $%
r=0 $ ($\underset{r\rightarrow 0}{\lim }R_{\mu \nu \lambda \kappa }R^{\mu
\nu \lambda \kappa }\rightarrow \infty $), and also the asymptotical
behavior of this spacetime is (anti)de Sitter ((a)dS) at $r\rightarrow
\infty $ ($\underset{r\rightarrow \infty }{\lim }R_{\mu \nu \lambda \kappa
}R^{\mu \nu \lambda \kappa }=\frac{8\Lambda ^{2}(\varepsilon )}{3}$). It is
worthwhile to mention that the asymptotical behavior of obtained solutions
is energy dependent. On the other hand, by replacing $f\left( \varepsilon
\right) =g\left( \varepsilon \right) =1$, the solution (\ref{BH}) reduces to
topological Reissner--Nordstr\"{o}m black hole solutions in (a)dS spacetime,
namely, $\psi \left( r\right) =k-\frac{m_{0}}{r}-\frac{\Lambda r^{2}}{3}+%
\frac{q^{2}}{r^{2}}$. In order to investigate the possibility of the
horizon, we study the behavior of obtained metric function (\ref{BH}) versus
$r$ in Fig. \ref{Fig1}. We found that the presented solutions may be
interpreted as black hole solutions with two horizons (inner and outer
horizons), extreme black hole (one horizon) or naked singularity (without
horizon).

\section{Thermodynamics}

In this section, we want to calculate the conserved and thermodynamic
quantities, and check the first law of thermodynamics for these black holes.

Considering the metric (\ref{metric}) and the obtained black hole solutions (%
\ref{BH}), the Hawking temperature of these black holes can be derived from
the definition of surface gravity at the outer horizon $r_{+} $ as follows
\cite{HawkingI,HawkingII}:
\begin{equation}
T=\frac{g\left( \varepsilon \right) }{4\pi f\left( \varepsilon \right) }%
\left. \psi {^{\prime }}\left( r,\varepsilon \right) \right\vert _{r=r_{+}}=%
\frac{1}{4\pi }\left( \frac{g\left( \varepsilon \right) \left[
kr_{+}^{2}-q^{2}(\varepsilon )f^{2}\left( \varepsilon \right) \right] }{%
f\left( \varepsilon \right) r_{+}^{3}}-\frac{\Lambda (\varepsilon )r_{+}}{%
f\left( \varepsilon \right) g\left( \varepsilon \right) }\right) .
\label{Temperature}
\end{equation}

It is notable that, the dependency on the rainbow functions indicates that
the temperature is modified.

The entropy of black holes satisfies the so-called area law of entropy in
general relativity. It means that the black hole's entropy equals to
one-quarter of horizon area (see Ref. \cite{Beckenstein}). Therefore, the
entropy of black holes is given by
\begin{equation}
S=\frac{r_{+}^{2}}{4g^{2}(\varepsilon )}.  \label{TotalS}
\end{equation}

The equation (\ref{TotalS}) shows that, the obtained entropy is modified in
gravity's rainbow.

In order to obtain the total charge of these solutions, one can employ the
Gauss law. Therefore, one can find the total electric charge in the
following form
\begin{equation}
Q=\frac{q(\varepsilon )f\left( \varepsilon \right) }{4\pi g\left(
\varepsilon \right) }.  \label{Q}
\end{equation}

The obtained total charge depends on rainbow functions. In other words, the
total charge is modified in this gravity.

In order to obtain the electric potential ($U$), we can calculate it on the
horizon with respect to a reference
\begin{equation}
U=A_{\mu }\chi ^{\mu }\left\vert _{r\rightarrow \infty }\right. -A_{\mu
}\chi ^{\mu }\left\vert _{r\rightarrow 0}\right. =\frac{q(\varepsilon )}{%
r_{+}}.  \label{U}
\end{equation}

Another important conserved quantity is related to total mass of the black
holes. For finding it, one can use Hamiltonian approach which results into
\begin{equation}
M=\frac{m_{0}(\varepsilon )}{8\pi f\left( \varepsilon \right) g\left(
\varepsilon \right) }.  \label{TotalM}
\end{equation}

Using the obtained conserved and thermodynamic quantities, we can check the
first law of thermodynamics for topological charged black hole solutions in
gravity's rainbow. For this purpose, we obtain the mass as a function of the
extensive quantities $S$ and $Q$. Considering Eqs. (\ref{BH}), (\ref{TotalS}%
) and (\ref{Q}) and by replacing them in Eq. (\ref{TotalM}), the mass $%
M(S,Q) $ is found as
\begin{equation}
M(S,Q)=\frac{4\Lambda (\varepsilon )S^{2}-3kS-12\pi ^{2}Q^{2}}{-12\pi
f\left( \varepsilon \right) \sqrt{S}}.
\end{equation}

Now, we define the intensive parameters conjugate to $S$ and $Q$. These
quantities are the temperature and the electric potential
\begin{eqnarray}
T &=&\left( \frac{\partial M(S,Q)}{\partial S}\right) _{Q}=\frac{4\Lambda
(\varepsilon )S^{2}-kS+4\pi ^{2}Q^{2}}{-8\pi f\left( \varepsilon \right)
S^{3/2}},  \label{T(S,Q)} \\
&&  \notag \\
U &=&\left( \frac{\partial M(S,Q)}{\partial Q}\right) _{S}=\frac{2\pi Q}{%
f\left( \varepsilon \right) \sqrt{S}},  \label{U(S,Q)}
\end{eqnarray}

The results of Eqs. (\ref{T(S,Q)}), (\ref{U(S,Q)}) coincide with Eqs. (\ref%
{Temperature}) and (\ref{U}), therefore, we find that these conserved and
thermodynamic quantities satisfy the first law of black hole thermodynamics
as
\begin{equation}
dM=TdS+UdQ.
\end{equation}

\section{Stability and geometrical thermodynamics}

Here, first, we are going to study the stability of solutions in the context
of heat capacity. Next, we consider the geometrical approach for studying
phase transitions. We investigate the effects of energy dependent parameters
on thermodynamical behavior and compare the results of both approaches.

In context of the canonical ensemble, the heat capacity is one of the
thermodynamical quantities carrying crucial information regarding thermal
structure of the black holes. The heat capacity includes three specific
interesting information. First, the discontinuities of this quantity mark
the possible thermal phase transitions that system can undergo. Second, the
sign of it determines whether the system is thermally stable or not. Indeed,
the positivity corresponds to thermal stability while the opposite shows
instability. Third, the roots of this quantity are also of interest since it
may yield the possible changes between stable/instable states or bound
point. Due to these important points, this section and the following one are
dedicated to calculation of the heat capacity of the solutions and
investigation of thermal structure of the black holes using such quantity.
We will show that by using this quantity alongside of the temperature, we
can draw a picture regarding the possible thermodynamical phase structures
of these black holes and stability/instability that these black holes could
enjoy/suffer. One can calculate the heat capacity in the following form
\begin{equation}
C_{Q}=T\left( \frac{\partial S}{\partial T}\right) _{Q}=\frac{\left( \frac{%
\partial M(S,Q)}{\partial S}\right) _{Q}}{\left( \frac{\partial ^{2}M(S,Q)}{%
\partial S^{2}}\right) _{Q}}=\frac{2S\left[ 4\Lambda (\varepsilon
)S^{2}-kS+4\pi ^{2}Q^{2}\right] }{4\Lambda (\varepsilon )S^{2}+kS-12\pi
^{2}Q^{2}}.
\end{equation}

In the context of black holes, it is argued that the root of heat capacity ($%
C_{Q}=T=0$) is representing a border line between physical ($T>0$) and
non-physical ($T<0$) black holes. We call it a physical limitation point.
The system in the case of this physical limitation point has a change in
sign of the heat capacity. In addition, it is believed that the divergencies
of the heat capacity represent phase transition critical points of black
holes. Therefore, the phase transition critical and limitation points of the
black holes in the context of the heat capacity are calculated with the
following relations
\begin{equation}
\left\{
\begin{array}{cc}
T=\left( \frac{\partial M(S,Q)}{\partial S}\right) _{Q}=0 & \text{%
physical~limitation~points} \\
\left( \frac{\partial ^{2}M(S,Q)}{\partial S^{2}}\right) _{Q}=0 & \text{%
phase~transition~critical points}%
\end{array}%
\right. .  \label{points}
\end{equation}

In order to find the physical limitation points, we consider Eq. (\ref%
{T(S,Q)}) and solve the following equation for the entropy:
\begin{equation}
T=\left( \frac{\partial M(S,Q)}{\partial S}\right) _{Q}=\frac{4\Lambda
(\varepsilon )S^{2}-kS+4\pi ^{2}Q^{2}}{-8\pi f\left( \varepsilon \right)
S^{3/2}}=0.
\end{equation}

We find two roots ($S_{1}$ and $S_{2}$) for the above equation as
\begin{equation}
\left\{
\begin{array}{c}
S_{1}=\frac{k-\sqrt{k^{2}-64\pi ^{2}\Lambda (\varepsilon )Q^{2}}}{8\Lambda
(\varepsilon )} \\
\\
S_{2}=\frac{k+\sqrt{k^{2}-64\pi ^{2}\Lambda (\varepsilon )Q^{2}}}{8\Lambda
(\varepsilon )}%
\end{array}%
\right. ,  \label{roots}
\end{equation}

In order to have real roots, we should consider $k^{2}-64\Lambda
(\varepsilon )\pi ^{2}Q^{2}\geq 0$, therefore, $Q\leq \pm \frac{k}{8\pi
\sqrt{\Lambda (\varepsilon )}}$. Our results about the physical limitation
points (\ref{roots}) show that these points depend on topological factor ($k$%
), the cosmological constant ($\Lambda (\varepsilon )$) and also the total
charge ($Q$). Considering the obtained limitation for the total charge ($%
Q\leq \pm \frac{k}{8\pi \sqrt{\Lambda (\varepsilon )}}$), we investigate the
behavior of the physical limitation points in table (I) and figures \ref%
{Fig2}, \ref{Fig3}, \ref{Fig4} and \ref{Fig5}.

\begin{table}[tbp]
\caption{The physical limitation points for $Q=0.02$.}
\label{tab1}
\begin{center}
\begin{tabular}{ccccc}
\hline\hline
$k $ & $\Lambda(\varepsilon )$ & $S_{1}$ & $S_{2}$ & number of points \\
\hline\hline
$1$ & $1$ & $0.017$ & $0.233$ & $2$ \\ \hline
$1$ & $-1$ & $0.015$ & $-$ & $1$ \\ \hline
$0$ & $1$ & $-$ & $-$ & $0$ \\ \hline
$0$ & $-1$ & $0.063$ & $-$ & $1$ \\ \hline
$-1$ & $1$ & $-$ & $-$ & $0$ \\ \hline
$-1$ & $-1$ & $0.265$ & $-$ & $1$ \\ \hline\hline
&  &  &  &
\end{tabular}%
\end{center}
\end{table}

Now, we investigate the obtained physical limitation points in table (I).
One should note it that the sign of temperature and heat capacity put a
restriction on the system as to it being physical or non-physical and also
stable or unstable. In other words, the negativity of temperature and heat
capacity is denoted as non-physical and unstable system, whereas the
positivity of temperature and heat capacity is denoted as a physical and
stable system. Here, our system is black hole.

\begin{figure}[tbp]
$%
\begin{array}{c}
\epsfxsize=8cm \epsffile{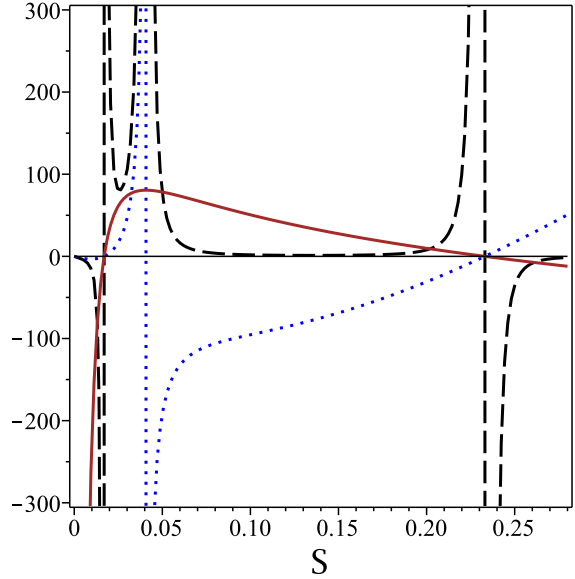}%
\end{array}
$%
\caption{$T$ (continuous line), $C_{Q}$ (dotted line) and $R_{HPEM}$ (dashed
line) versus $S$ for $Q=0.02$, $\Lambda (\protect\varepsilon )=1$, $f(%
\protect\varepsilon)=g(\protect\varepsilon)=1.1$, $k=1$.}
\label{Fig2}
\end{figure}

\begin{figure}[tbp]
$%
\begin{array}{cc}
\epsfxsize=7cm \epsffile{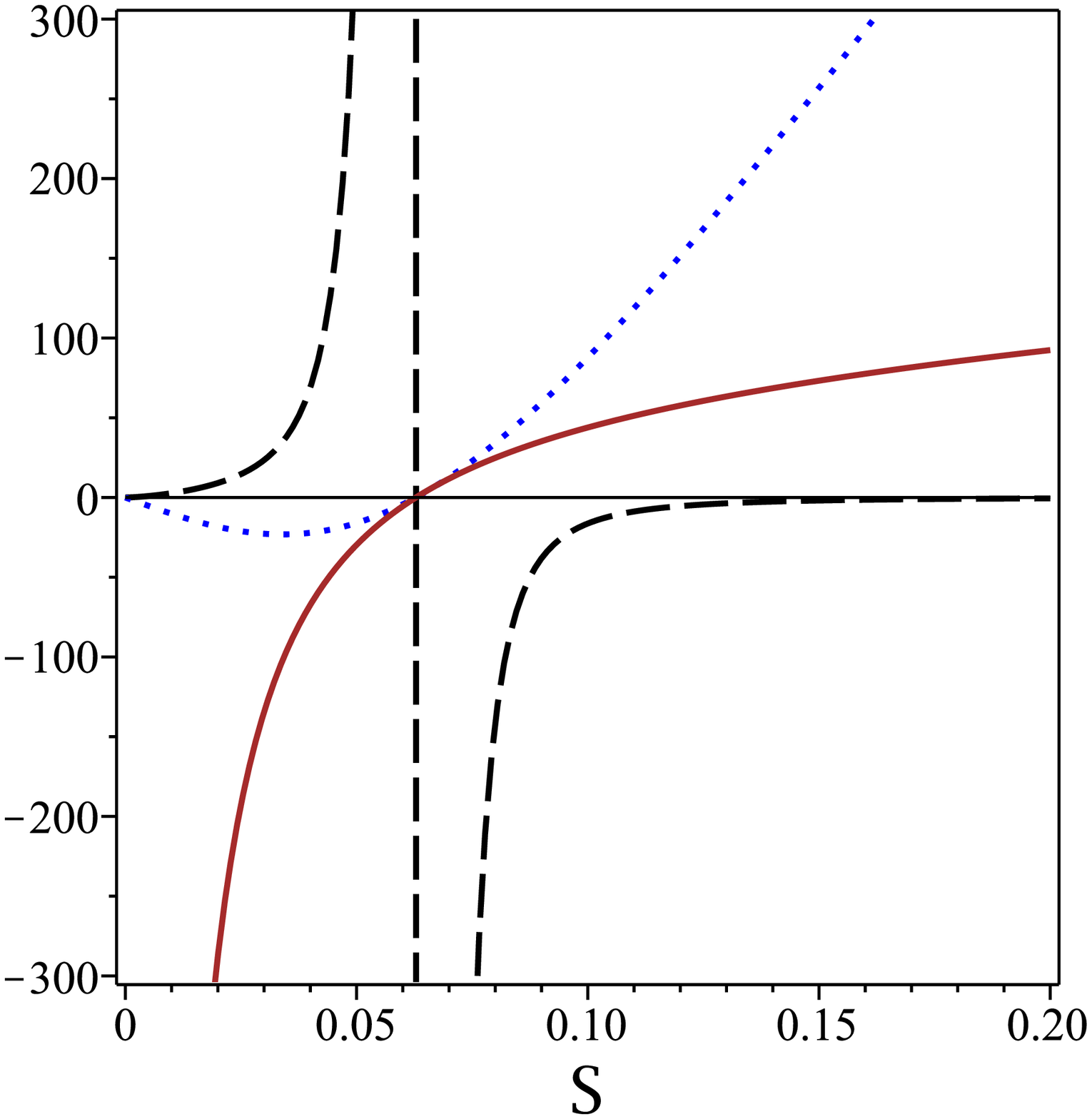} & \epsfxsize=7cm \epsffile{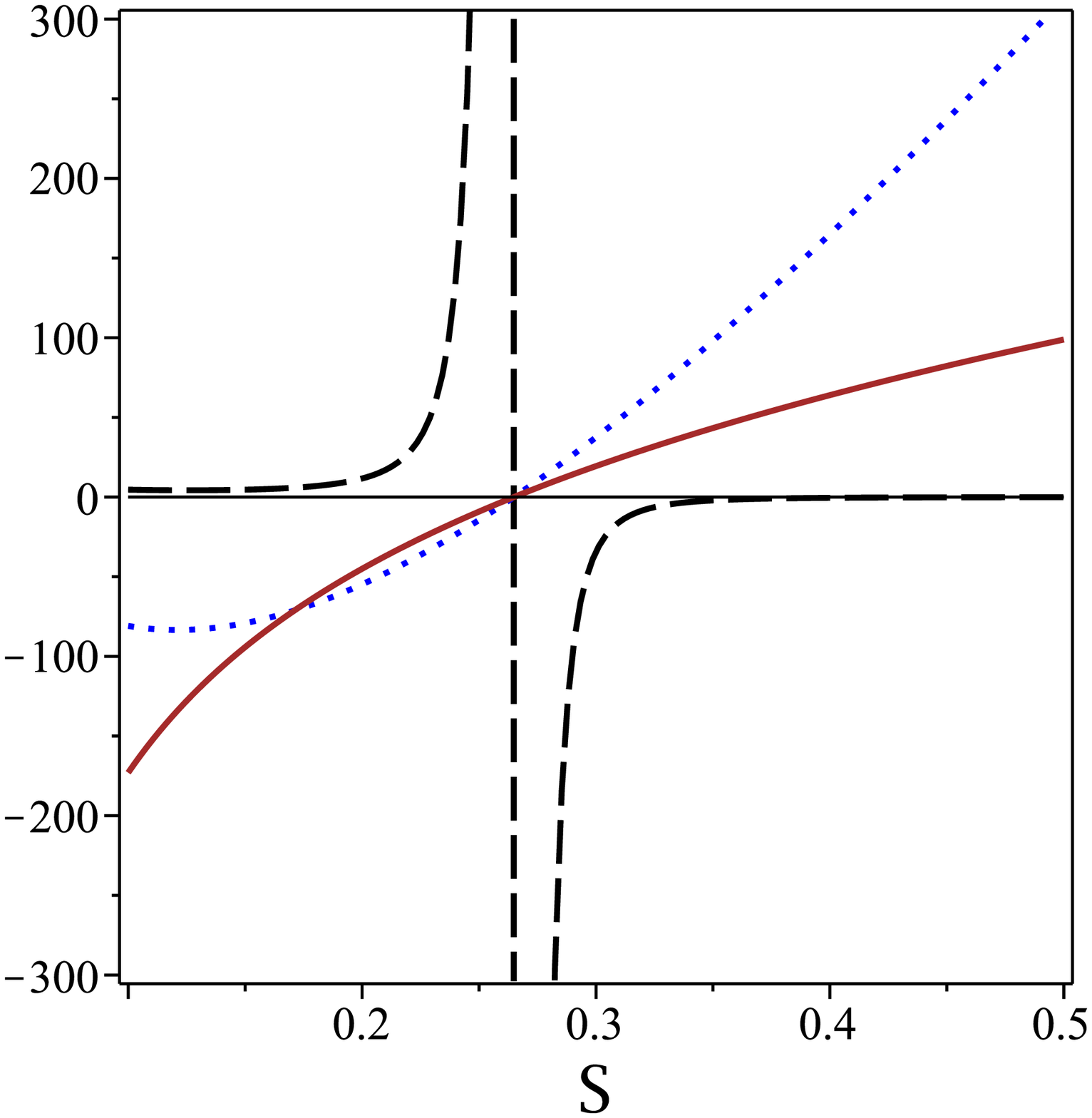}%
\end{array}
$%
\caption{$T$(continuous line), $C_{Q}$ (dotted line) and $R_{HPEM}$ (dashed
line) versus $S$ for $Q=0.02$, $\Lambda (\protect\varepsilon )=-1$, $f(%
\protect\varepsilon)=g(\protect\varepsilon)=1.1$. \newline
\textbf{Left diagram:} for $k=0$. \textbf{Right diagram:} for $k=-1$.}
\label{Fig3}
\end{figure}

\begin{figure}[tbp]
$%
\begin{array}{c}
\epsfxsize=8cm \epsffile{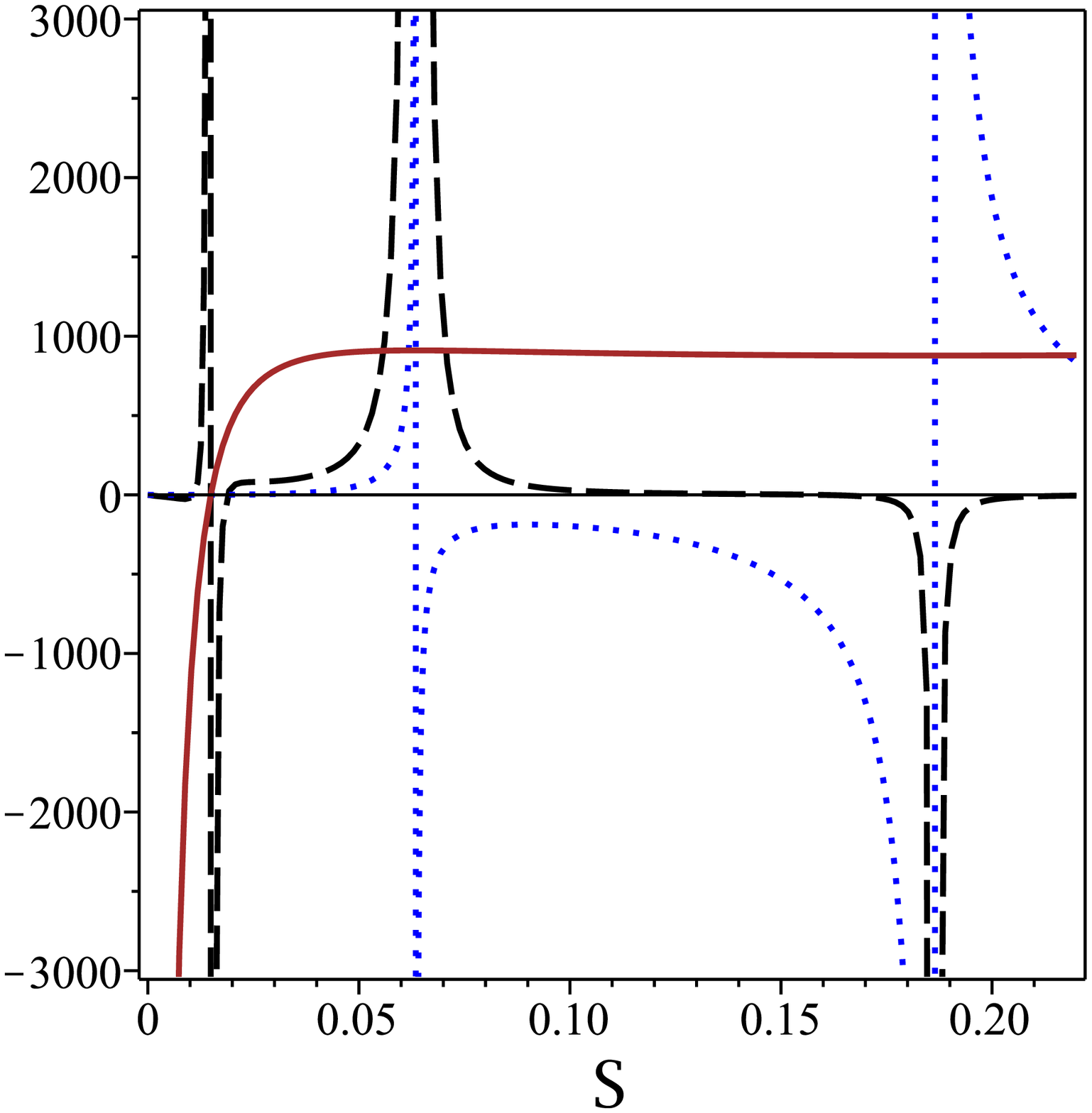}%
\end{array}
$%
\caption{$T$ (continuous line), $C_{Q}$ (dotted line) and $R_{HPEM}$ (dashed
line) versus $S$ for $Q=0.02$, $\Lambda (\protect\varepsilon )=1$, $f(%
\protect\varepsilon)=g(\protect\varepsilon)=1.1$, $k=1$.}
\label{Fig4}
\end{figure}

\begin{figure}[tbp]
$%
\begin{array}{cc}
\epsfxsize=7cm \epsffile{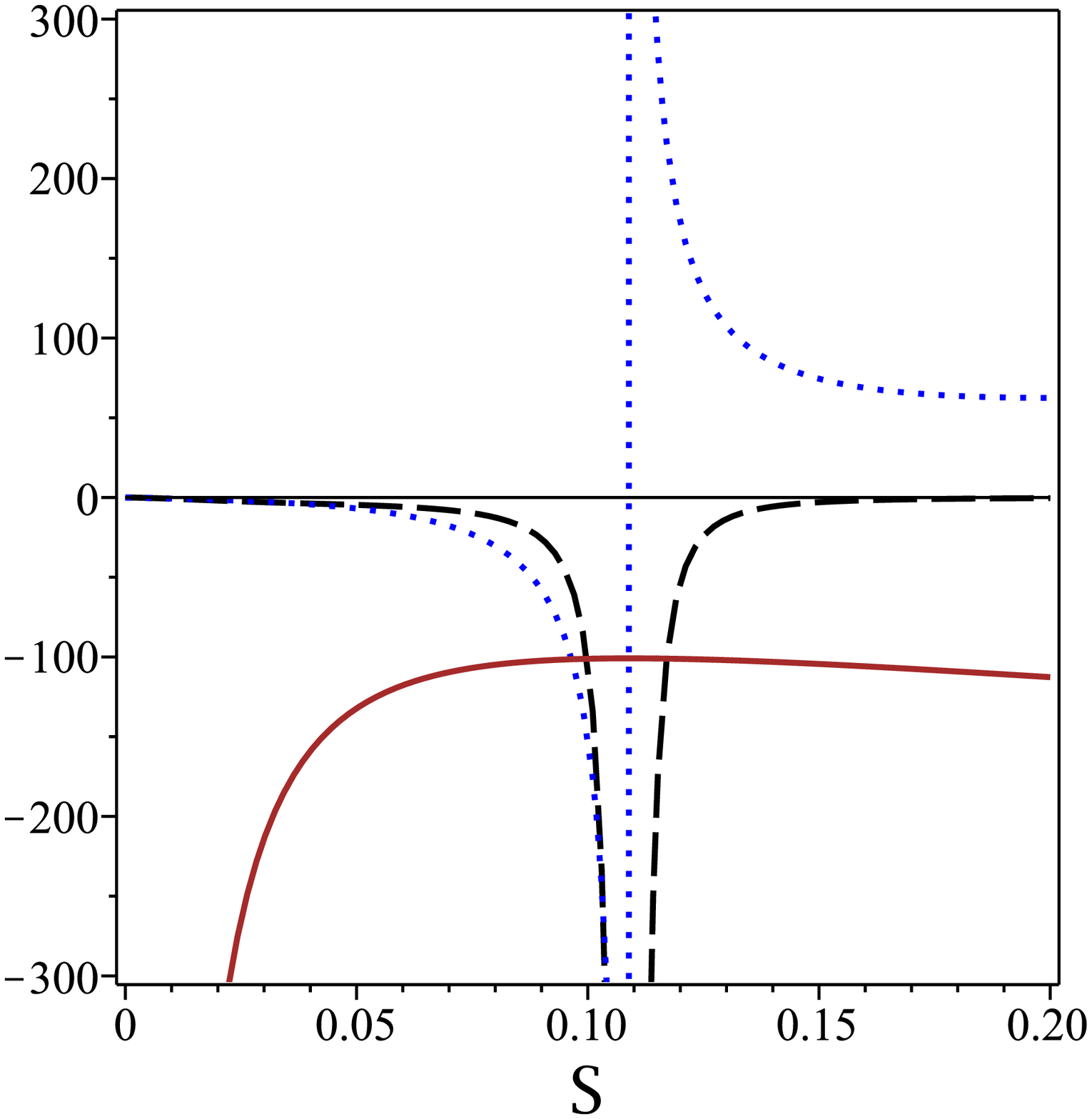} & \epsfxsize=7cm \epsffile{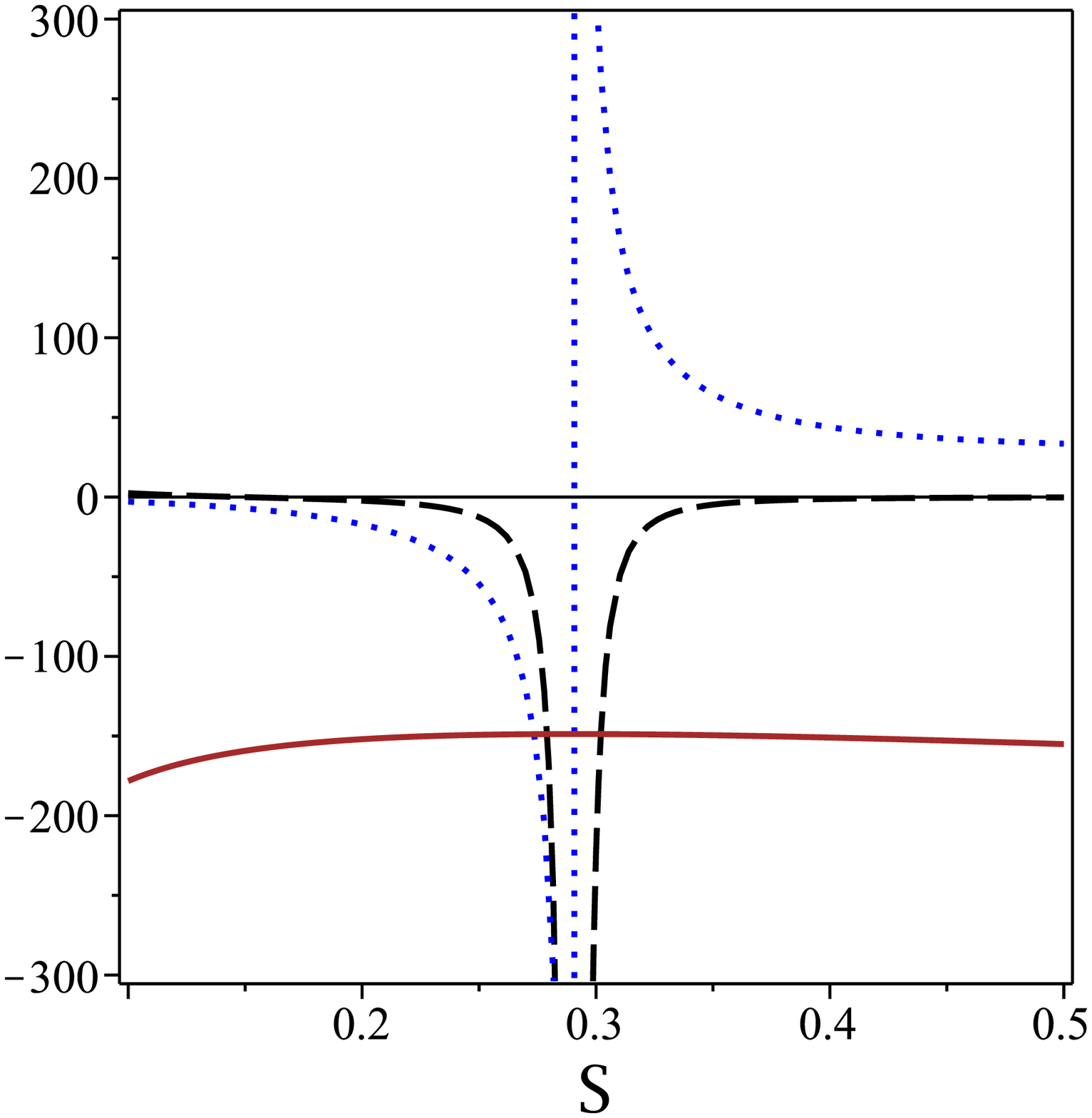}%
\end{array}
$%
\caption{$T$(continuous line), $C_{Q}$ (dotted line) and $R_{HPEM}$ (dashed
line) versus $S$ for $Q=0.02$, $\Lambda (\protect\varepsilon )=1$, $f(%
\protect\varepsilon)=g(\protect\varepsilon)=1.1$. \newline
\textbf{Left diagram:} for $k=0$. \textbf{Right diagram:} for $k=-1$.}
\label{Fig5}
\end{figure}

Our solutions include three different cases for topological charged black
holes in gravity's rainbow:

\textit{Case I}: existence of two roots for the temperature and the heat
capacity (Fig. \ref{Fig2}). In this case, there are two critical entropy,
namely $S_{1}$ and $S_{2}$, in which for $S<S_{1}$, the temperature and the
heat capacity of black holes are negative. Therefore, the black holes with
small entropy are non-physical and unstable. For $S_{2}<S$, the heat
capacity is positive but the temperature is negative. Therefore the black
holes with large entropy are non-physical. It is notable that, there is an
important region which is related to $S_{1}<S<S_{{div}}$, where $S_{{div}}$
is divergence point. Temperature and heat capacity of black holes in this
region are positive. In other words, the obtained black holes are physical
and enjoy thermal stability when their entropy is in range $S_{1}<S<S_{{div}%
} $. Also, for region $S_{{div}}<S<S_{2}$, the temperature is positive but
the heat capacity is negative. Indeed, the black holes are physical but
unstable (see Fig. \ref{Fig2}).

\textit{Case II}: there is a critical entropy, $S_{1}$. Here, we have two
different behaviors. First: for $S<S_{1}$, temperature and heat capacity of
the black holes are negative and therefore our systems are non-physical and
unstable. But for $S>S_{1}$, these quantities are positive and therefore our
black holes are physical and stable. In other words, our results confirm
that the black holes with large entropy are physical and enjoy thermal
stability but for small entropy the black holes are non-physical and
unstable (see Fig. \ref{Fig3}). Second: the temperature and the heat
capacity are negative and so the systems are non-physical and unstable in
the range $S<S_{1}$, but for $S>S_{1}$, the temperature is positive whiles
the heat capacity has different sign (see Fig. \ref{Fig4}).

\textit{Case III}: there is no root. In this case, the temperature of black
holes is negative. Although the heat capacity of these black holes may be
positive but the temperature of black holes is always negative. In other
words, the obtained black holes are non-physical (see Fig. \ref{Fig5}).

In order to study the phase transition critical points, we obtain the
following relation
\begin{equation}
\left( \frac{\partial ^{2}M(S,Q)}{\partial S^{2}}\right) _{Q}=4\Lambda
(\varepsilon )S^{2}+kS-12\pi ^{2}Q^{2}=0.
\end{equation}

Using the above equation, we obtain two divergence points for heat capacity (%
$S_{1}^{\ast }$ and $S_{2}^{\ast }$) as
\begin{equation}
\left\{
\begin{array}{c}
S_{1}^{\ast }=\frac{-k+\sqrt{k^{2}+192\pi ^{2}\Lambda (\varepsilon )Q^{2}}}{%
8\Lambda (\varepsilon )} \\
\\
S_{2}^{\ast }=\frac{-k-\sqrt{k^{2}+192\pi ^{2}\Lambda (\varepsilon )Q^{2}}}{%
8\Lambda (\varepsilon )}%
\end{array}%
\right. ,
\end{equation}

As one can see, our results show that, the phase transition critical points
depend on topological factor ($k$), the cosmological constant ($\Lambda
(\varepsilon )$) and also the total charge ($Q$). Here, we are going to
investigate the behavior of the phase transition critical points in table
(II) and figures \ref{Fig2}-\ref{Fig5}.

\begin{table}[tbp]
\caption{The phase transition critical points for $Q=0.02$.}
\label{tab2}
\begin{center}
\begin{tabular}{ccccc}
\hline\hline
$k $ & $\Lambda(\varepsilon )$ & $S_{1}^{\ast }$ & $S_{2}^{\ast }$ & number
of points \\ \hline\hline
$1$ & $1$ & $0.041$ & $-$ & $1$ \\ \hline
$1$ & $-1$ & $0.063$ & $0.186$ & $2$ \\ \hline
$0$ & $1$ & $0.109$ & $-$ & $1$ \\ \hline
$0$ & $-1$ & $-$ & $-$ & $0$ \\ \hline
$-1$ & $1$ & $0.291$ & $-$ & $1$ \\ \hline
$-1$ & $-1$ & $-$ & $-$ & $0$ \\ \hline\hline
&  &  &  &
\end{tabular}%
\end{center}
\end{table}

According to the obtained results for phase transition critical points (or
divergence points) in table (II) and figures \ref{Fig2}-\ref{Fig5}, we may
encounter with three different cases:

\textit{Case I}: there are two divergencies for the heat capacity ($%
S_{1}^{\ast }$ and $S_{2}^{\ast }$). Since the black holes are physical and
unstable between two divergencies (the temperature is positive but the heat
capacity is negative), therefore, a phase transition from the black holes
with small entropy (small black holes) to large entropy (large black holes)
occurs between two divergencies (see Fig. \ref{Fig4}). This phase transition
is similar to the Van der Waals phase transition for black hole. It has been
found that the thermodynamics of an asymptotically AdS metric in $4$%
-dimensional spacetime matches exactly with the thermodynamics of the Van
der Waals fluid. In other words, the isocharge in the temperature-entropy
plane has an unstable branch and two stable ones when the charge below a
critical value, and also there exists a second-order critical point at a
critical charge. Therefore, for $k=1$, we encounter with a Van der\
Waals-like phase transition of black hole in gravity's rainbow (see Fig. \ref%
{Fig4}). Recently, the research on a Van der Waals-like phase transition has
been generalized to the extended phase space \cite%
{VdWI,VDWI2,VdWII,VdWIII,VdWIV,VdWV,VdWVI}. In this framework, the
cosmological constant is taken as a thermodynamical pressure ($P=-\frac{%
\Lambda }{8\pi }$), and its conjugate quantity is treated as the
thermodynamical volume \cite{Therm1}. Considering this view, there is a
precise pressure-volume oscillatory behavior and the small-large black hole
phase transition is identified with the liquid-gas phase transition of the
van der Waals fluid \cite{KM}. This feature has been investigated for black
holes with more details in refs. \cite{PTI,PTII,PTIII,PTIV,PTV,PTVI,PTVII}.

\textit{CaseII}: existence of one divergency ($S_{1}^{\ast }$). In this case
we encounter with two different cases which depend on the sign of
temperature of black holes. First: there is a phase transition from the
unstable black holes with large entropy to the stable black holes with small
entropy (see Fig. \ref{Fig2}). Second: there is no phase transition when the
temperature is negative because the black holes are non-physical (see Fig. %
\ref{Fig5}).

\textit{CaseIII}: there is no divergence point. It is notable that, the
maximum of temperature is where the system acquires divergency in its heat
capacity. Therefore, the absence of divergency in the heat capacity is due
to the fact that there is no the maximum for the temperature of black holes
(see Fig. \ref{Fig3}).

Another approach for studying the phase transition critical points of black
holes is related to geometrical thermodynamics. There are several metrics
that one can employ in order to build a geometrical phase space by
thermodynamical quantities such as Weinhold \cite{WeinholdI,WeinholdII},
Ruppeiner \cite{RuppeinerI,RuppeinerII}, Quevedo \cite{QuevedoI,QuevedoII},
and HPEM \cite{HPEM}. It was previously argued that Ruppeiner, Weinhold and
Quevedo metrics may not provide us with a completely flawless mechanism for
studying the geometrical thermodynamics of specific types of black holes
(see Refs. \cite{HPEMI,HPEMII,HPEMIV}, for more details). Therefore, in this
paper, we consider the HEPM metric and investigate geometrical
thermodynamics of topological charged balck holes in gravity's rainbow.

The HPEM metric is given by \cite{HPEM}
\begin{equation}
dS_{HPEM}^{2}=\frac{SM_{S}}{M_{QQ}^{3}}\left(
-M_{SS}dS^{2}+M_{QQ}dQ^{2}\right) ,  \label{HPEM}
\end{equation}%
where $M_{S}=$ $\left( \frac{\partial M(S,Q)}{\partial S}\right) _{Q}$, $%
M_{SS}=\left( \frac{\partial ^{2}M(S,Q)}{\partial S^{2}}\right) _{Q}$ and $%
M_{QQ}=\left( \frac{\partial ^{2}M(S,Q)}{\partial Q^{2}}\right) _{S}$.
Calculations show that the numerator and denominator of Ricci scalar of HPEM
metric (\ref{HPEM}), are given by the following forms
\begin{eqnarray}
\text{numerator }(R_{HPEM}) &=&S^{2}M_{S}^{2}M_{QQQ}^{3}M_{SSS}\left( \frac{%
M_{SS}}{M_{S}}-\frac{1}{S}\right) +S^{2}M_{S}^{2}M_{QQ}^{3}M_{SQQ}\left(
2M_{SSS}+\frac{M_{SS}}{S}-\frac{M_{SS}^{2}}{M_{S}}\right)  \notag \\
&&+S^{2}M_{S}^{2}M_{SS}^{2}M_{QQQ}^{2}\left( \frac{M_{SQ}M_{QQ}}{M_{S}M_{QQQ}%
}-9\right) +6S^{2}M_{S}^{2}M_{QQ}^{2}M_{SS}^{2}\left( \frac{M_{QQQQ}}{M_{QQ}}%
-\frac{M_{SSQQ}}{M_{SS}}\right)  \notag \\
&&+S^{2}M_{SQ}^{2}M_{SS}^{2}M_{QQ}^{2}\left( 2-\frac{M_{S}M_{SSQ}}{%
M_{SS}M_{SQ}}\right) +S^{2}M_{QQ}^{2}\left(
M_{S}^{2}M_{SSQ}^{2}-2M_{SS}^{3}M_{QQ}\right)  \notag \\
&&+S^{2}M_{S}^{2}M_{SS}M_{QQ}\left( 2M_{SQQ}^{2}+4M_{QQQ}M_{SSQ}\right)
-2M_{S}^{2}M_{SS}M_{QQ}^{3},  \label{num} \\
&&  \notag \\
\text{denominator }(R_{HPEM}) &=&2S^{3}M_{S}^{3}M_{SS}^{2},  \label{dnom}
\end{eqnarray}%
where $M_{XX}=\left( \frac{\partial ^{2}M}{\partial X^{2}}\right) $, $%
M_{XY}=\left( \frac{\partial ^{2}M}{\partial X\partial Y}\right) $, $%
M_{XXX}=\left( \frac{\partial ^{3}M}{\partial X^{3}}\right) $, $%
M_{XXXX}=\left( \frac{\partial ^{4}M}{\partial X^{4}}\right) $ and also $%
M_{XXYY}=\left( \frac{\partial ^{4}M}{\partial X^{2}\partial Y^{2}}\right) $.

Our results show that, denominator of the Ricci scalar of the HPEM metric
contains numerator and denominator of the heat capacity. In other words,
divergence points of the Ricci scalar of HPEM metric coincide with both
roots (the physical limitation points) and phase transition critical points
of the heat capacity. Therefore, all the physical limitation and the phase
transition critical points are included in the divergencies of the Ricci
scalar of HPEM metric (see Figs. \ref{Fig2}-\ref{Fig5}). Another important
results of HPEM metric is related to the different behavior of Ricci scalar
before and after its divergence points. It was seen that the behavior of
Ricci scalar for divergence points related to the physical limitation and
phase transition critical points is different. In other words, the sign of
Ricci scalar change before and after divergencies when the heat capacity is
zero (see figs. \ref{Fig2}-\ref{Fig4}). But the signs of Ricci scalar are
the same when the heat capacity encounter with divergencies (see figs. \ref%
{Fig2}, \ref{Fig4} and \ref{Fig5}). These divergencies called $\Lambda $
divergencies. Therefore, considering this approach also enable us to
distinguish physical limitation and phase transition critical points from
one another.

\section{Heat efficiency}

\begin{figure}[tbp]
\centering
$%
\begin{array}{c}
\epsfxsize=15cm \epsffile{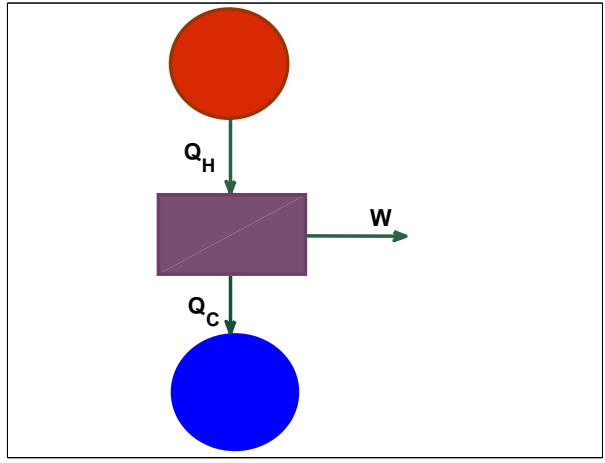}%
\end{array}
$%
\caption{The heat engine flows.}
\label{Fig6}
\end{figure}

In this section, we are going to obtain another important quantity for black
holes which is called the heat efficiency. In this paper, we are interested
in classical heat engine. A heat engine is a physical system that takes heat
from warm reservoir and turns a part of it into the work while its remaining
is given to cold reservoir (see Fig. (\ref{Fig6}), for more details). In
order to calculate work done by the heat engine and given the equation of
state, one can use the $P-V$ \ diagrams for describing the heat engine which
results into a closed path \cite{Johnson,JohnsonI,JohnsonIa}.\ On the other
hand, Wei and Liu showed the work or heat through measuring the areas by
casting the process into the $T-S$\ diagram, see Fig. $2$\ in ref. \cite%
{WeiHeat} for more details. It is notable that, for a thermodynamics cycle,
one may extract mechanical work via the $PdV$ term in the first law of
thermodynamics as $W=Q_{H}-Q_{C}$ (the first law of thermodynamics is given
as $\triangle U=\triangle Q-W$). According to this fact that $\triangle U$
(the internal energy changes) is zero for a thermodynamic cycle, the first
law of thermodynamics reduces to $W=\triangle Q=Q_{H}-Q_{C}$, where $Q_{H}$
is a net input heat flow, $Q_{C}$ is a net output flow and also $W$ is a net
output work. On the other hand, the efficiency of heat engine is defined as%
\begin{equation}
\eta =\frac{W}{Q_{H}}=\frac{Q_{H}-Q_{C}}{Q_{H}}=1-\frac{Q_{C}}{Q_{H}},
\label{efficiency}
\end{equation}


The heat engine depends on the choice of path in the $P-V$ diagram and also
the equation of state of the black hole in question. Some classical cycles
such as Carnot cycle includes a pair of isotherms at temperatures $T_{H}$
and $T_{C}$ where $T_{H}>T_{C}$ ($T_{H}$ and $T_{C}$ are temperatures of the
warm and the cold reservoirs, respectively), for this cycle, there is a pair
of isotherms with different temperatures where cycle has maximum efficiency
and it is described by $\eta =1-\frac{T_{C}}{T_{H}}$. It is noteworthy that
for this case, there is an isothermal expansion where the system absorbs
some heat and an isothermal compression during expulsion of some heat of the
system. It is possible to show that these two paths of isotherms are
connected to each other by different methods. The first method is using
isochoric path, like classical Stirling cycle. The second one is adiabatic
path similar to classical Carnot cycle. Therefore, the form of path is a key
factor for the definition of cycle.

It is notable that the thermodynamic volume of topological black holes in
gravity's rainbow is given as
\begin{equation}
V=\frac{r_{+}^{3}}{3f\left( \varepsilon \right) g^{3}\left( \varepsilon
\right) }.  \label{V}
\end{equation}%
\textbf{\ }

On the other hand, for the obtained black holes in this gravity, the entropy
$S$ and the thermodynamic volume $V$\ are related by Eqs. (\ref{TotalS}) and
(\ref{V}) as $S=\frac{r_{+}^{2}}{4g^{2}\left( \varepsilon \right) }=\frac{%
\left[ 3Vf\left( \varepsilon \right) \right] ^{2/3}}{4}$.\ In other words,
entropy depends on volume ($S\propto V$). It means that adiabatic and
isochores are the same (Carnot and Stirling methods coincide with each
other). Therefore, the efficiency of cycle can be calculated easily.

\begin{figure}[tbp]
\centering
$%
\begin{array}{c}
\epsfxsize=15cm \epsffile{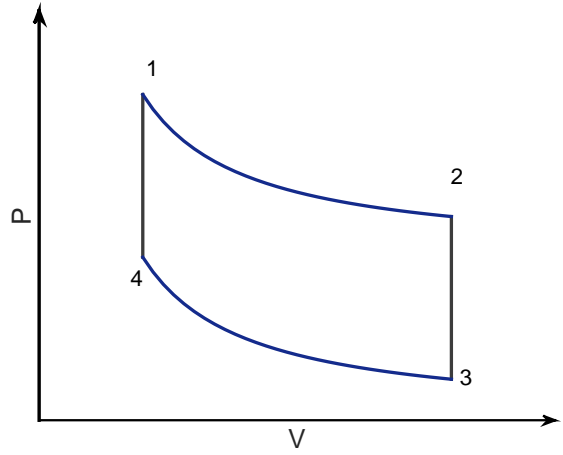}%
\end{array}
$%
\caption{Carnot cycle.}
\label{Fig7}
\end{figure}

So along the upper isotherm (Fig. (\ref{Fig7})) and by considering Eqs. (\ref%
{TotalS}) and (\ref{V}), the net input heat flow is given as%
\begin{equation}
Q_{H}=T_{H}\Delta S_{1\rightarrow 2}=\frac{T_{H}}{4}\left( 3f\left(
\varepsilon \right) \right) ^{2/3}\left( V_{2}^{2/3}-V_{1}^{2/3}\right) ,
\end{equation}%
and also along the lower isotherm (Fig. (\ref{Fig7})) and by considering
Eqs. (\ref{TotalS}) and (\ref{V}), the net output heat flow is given as
\begin{equation}
Q_{C}=T_{C}\Delta S_{3\rightarrow 4}=\frac{T_{C}}{4}\left( 3f\left(
\varepsilon \right) \right) ^{2/3}\left( V_{3}^{2/3}-V_{4}^{2/3}\right) ,
\end{equation}%
which according to Fig. (\ref{Fig7}), we have $V_{1}=V_{4}$ and $V_{2}=V_{3}$%
, and by using the equation (\ref{efficiency}), the efficiency of heat
engine becomes
\begin{equation}
\eta =1-\frac{Q_{C}}{Q_{H}}=1-\frac{T_{C}}{T_{H}}.
\end{equation}

It is notable that, the heat engine for black holes was proposed by Johnson
in 2014 \cite{Johnson}. Considering the concepts introduced by Johnson, the
heat engines provided by black holes in various gravities have been
investigated in some literatures. For example; the heat engine for
Gauss-Bonnet \cite{JohnsonI}, Born-Infeld AdS \cite{JohnsonII}, dilatonic
Born-Infeld \cite{Bhamidipati}, charged AdS black holes \cite{ChargeHeat},
Black hole in conformal gravity \cite{ConformalHeat}, Kerr AdS and dyonic
black holes \cite{Sadeghi}, BTZ \cite{Mo}, polytropic black holes \cite%
{Setare}, black holes in massive gravity \cite{Hendiheat} and Accelerating
AdS black holes \cite{Accelerating} have been studied (see refs. \cite%
{HeatmoreI,HeatmoreII,HeatmoreIII,HeatmoreIV}, for more details). Also, Wei
and Liu suggested that the working substance is the fluid constituted with
the virtual black hole molecules which carry the degree of freedom of black
hole entropy \cite{WeiLiu}. Then they studied the black hole heat engine by
a charged anti de Sitter black hole. In the reduced $T-S$\ chart, it was
found that the work, heat and efficiency of the engine are independent of
the black hole charge. Also, their results showed that the black hole engine
working\ along the Rankine cycle with a back pressure mechanism has a higher
efficiency. Their results provided efficient mechanism to produce the useful
mechanical work with black hole and such heat engine may act as a possible
energy source for the high energy astrophysical phenomena near the black
hole \cite{WeiHeat}.

Here, we want to obtain the efficiency of heat engine for topological black
holes in gravity's rainbow. For this purpose, by inserting $\Lambda
(\varepsilon )=-8\pi P$ in the obtained temperature of black holes (Eq. (\ref%
{Temperature})), we have
\begin{equation}
T=\frac{1}{4\pi }\left( \frac{kg\left( \varepsilon \right) }{f\left(
\varepsilon \right) r_{+}}+\frac{8\pi Pr_{+}}{f\left( \varepsilon \right)
g\left( \varepsilon \right) }\right) .  \label{TP}
\end{equation}

Solving the equation (\ref{TP}), we obtain the pressure in the following
form
\begin{equation}
P=\frac{g\left( \varepsilon \right) }{8\pi r_{+}^{2}}\left( 4\pi
r_{+}f\left( \varepsilon \right) T-kg\left( \varepsilon \right) \right) .
\end{equation}

There are two different heat capacities for a system; the heat capacity at
constant pressure ($C_{P}$) and the heat capacity at constant volume ($C_{V}$%
). The heat capacities at constant pressure and at constant volume can be
calculated by the standard thermodynamic relations, as
\begin{eqnarray}
C_{V} &=&T\frac{\partial S}{\partial T}\left\vert _{V}\right. ,  \label{Cv}
\\
&&  \notag \\
C_{P} &=&T\frac{\partial S}{\partial T}\left\vert _{P}\right. .  \label{Cp}
\end{eqnarray}

\begin{figure}[tbp]
$%
\begin{array}{c}
\epsfxsize=15cm \epsffile{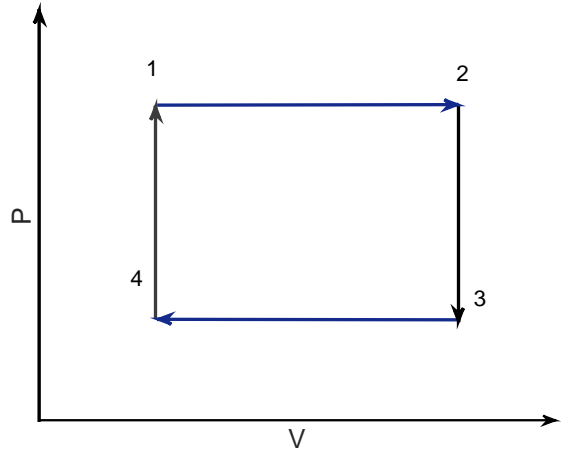}%
\end{array}
$%
\caption{$P$-$V$ diagram.}
\label{Fig8a}
\end{figure}

According to this fact that the entropy of black holes in gravity's rainbow
is a regular function of the thermodynamic volume $V$ ($S\propto V$), the
heat capacity at constant volume will vanish, $C_{V}=0$. Therefore an
explicit expression for $C_{P}$ would suggest that there should be a new
engine which includes two isobars and two isochores/adiabatic similar to
Fig. \ref{Fig8a}.

For this purpose, we can consider a rectangle cycle in the $P-V$ plane (Fig. %
\ref{Fig8a}) which consists two isobars (paths of $1\rightarrow 2$ and $%
3\rightarrow 4$) and two isochores (paths of $2\rightarrow 3$ and $%
4\rightarrow 1$). Also, a possible scheme for this heat engine involves
specifying values of temperature where $T_{2}=T_{H}$ and $T_{4}=T_{C}$.
According to the fact that the paths of $1\rightarrow 2$ and $3\rightarrow 4$
are isobars, we find $P_{1}=P_{2}$ and $P_{3}=P_{4}$. So, we can calculate
the work which is done in this cycle as
\begin{eqnarray}
W &=&\oint PdV=W_{1\rightarrow 2}+W_{2\rightarrow 3}+W_{3\rightarrow
4}+W_{4\rightarrow 1}  \notag \\
&=&W_{1\rightarrow 2}+W_{3\rightarrow 4}=\left( P_{1}-P_{4}\right) \left(
V_{2}-V_{1}\right) .  \label{WorkI}
\end{eqnarray}

It is notable that the works which are done in the paths of $2\rightarrow 3$
and $4\rightarrow 1$ are isochores, therefore, these terms are zero.

On the other hand, $C_{P}$ for these black holes is calculated as
\begin{equation}
C_{P}=T\frac{\partial S}{\partial T}\left\vert _{P}\right. =T\frac{\left(
\frac{\partial S}{\partial r_{+}}\right) }{\left( \frac{\partial T}{\partial
r_{+}}\right) }=\frac{\left( 8\pi r_{+}^{2}f\left( \varepsilon \right)
T+kg^{2}\left( \varepsilon \right) \right) r_{+}^{2}}{2g^{2}\left(
\varepsilon \right) \left( 8\pi r_{+}^{2}P-kg^{2}\left( \varepsilon \right)
\right) }.  \label{Cpp}
\end{equation}

Here, we want to consider the large temperature limit ($T>>0$). In other
words, we can discard the terms in which the temperature are in the
denominator, i. e. $\frac{1}{T},$ $\frac{1}{T^{2}}$, ... have small values.
In this paper, we keep the term, $\frac{1}{T}$, and omit higher orders of $T$
($\frac{1}{T^{2}}$, ... ) in order to extract $r_{+}$ from the equation (\ref%
{TP}) resulting into
\begin{equation}
r_{+}=\frac{f\left( \varepsilon \right) g\left( \varepsilon \right) T}{2P}-%
\frac{kg\left( \varepsilon \right) }{4\pi f\left( \varepsilon \right) T}+...,
\end{equation}%
which by inserting the above equation in the thermodynamic volume (\ref{V})
and the heat capacity (\ref{Cpp}), we obtain $V$ and $C_{P}$ as
\begin{eqnarray}
V &=&\frac{r_{+}^{3}}{3f\left( \varepsilon \right) g^{3}\left( \varepsilon
\right) }=\frac{f^{2}\left( \varepsilon \right) T^{3}}{24P^{3}}-\frac{kT}{%
16\pi P^{2}}+...~, \\
&&  \notag \\
C_{P} &=&\frac{f^{2}\left( \varepsilon \right) T^{2}}{8P^{2}}+...~.
\end{eqnarray}

Now, we are in a position to obtain the heat input and the work done by the
heat engine
\begin{eqnarray}
Q_{H} &=&\int_{T_{1}}^{T_{2}}C_{p}dT=\frac{f^{2}\left( \varepsilon \right) }{%
24P_{1}^{2}}\left( T_{2}^{3}-T_{1}^{3}\right) +...~, \\
&&  \notag \\
W &=&\left( P_{1}-P_{4}\right) \left( V_{2}-V_{1}\right) =\left( 1-\frac{%
P_{4}}{P_{1}}\right) \left( \frac{f^{2}\left( \varepsilon \right) \left(
T_{2}^{3}-T_{1}^{3}\right) }{24P_{1}^{2}}-\frac{k\left( T_{2}-T_{1}\right) }{%
16\pi P_{1}}+...\right).
\end{eqnarray}

As one can see, the heat input depends on rainbow function ($f^{2}\left(
\varepsilon \right) $). Also, the work done by the heat engine ($W$) depends
on both rainbow function ($f^{2}\left( \varepsilon \right) $) and
topological factor ($k$).

After some calculation, the engine efficiency is given as
\begin{equation}
\eta =\frac{W}{Q_{H}}=\left( 1-\frac{P_{4}}{P_{1}}\right) \left[ 1-\frac{%
3kP_{1}}{2\pi f^{2}\left( \varepsilon \right) \left(
T_{1}^{2}+T_{1}T_{2}+T_{2}^{2}\right) }+...\right] .  \label{eta}
\end{equation}

In order to have positive efficiency and this fact that the efficiency
satisfies the relation $\eta \leq 1$, we obtain the following condition for
topological factor as
\begin{equation}
k<\frac{2\pi f^{2}\left( \varepsilon \right) \left(
T_{1}^{2}+T_{1}T_{2}+T_{2}^{2}\right) }{3P_{1}}.
\end{equation}

\begin{figure}[tbp]
$%
\begin{array}{c}
\epsfxsize=7cm \epsffile{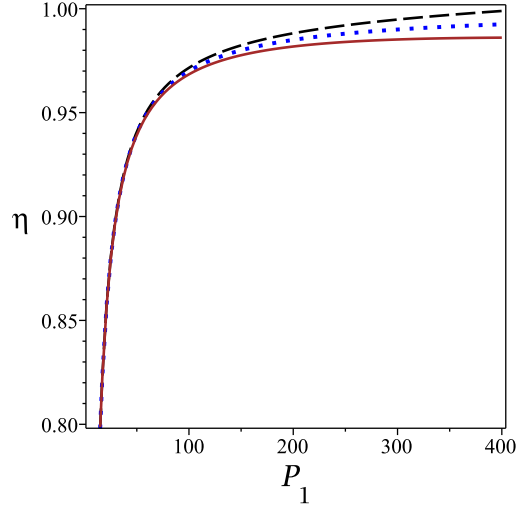}%
\end{array}
$%
\caption{$\protect\eta $ versus $P_{1}$ for $T_{2}=100$, $T_{1}=80$, $%
P_{4}=3 $, $f(\protect\epsilon )=1.1$, $k=-1$ (dashed line), $k=0$ (dotted
line) and $k=1$ (continuous line).}
\label{Fig9}
\end{figure}
\begin{figure}[tbp]
$%
\begin{array}{c}
\epsfxsize=7cm \epsffile{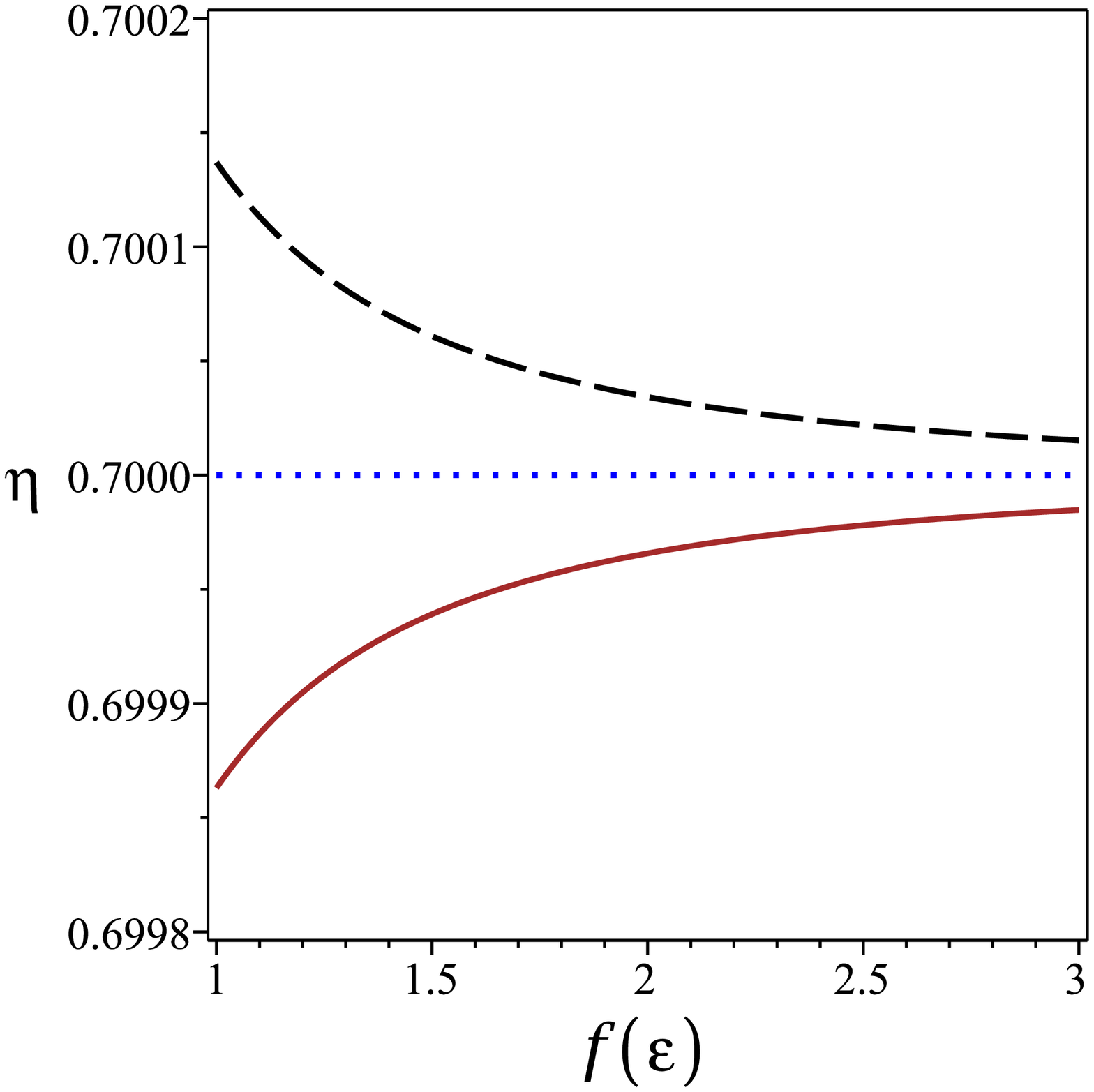}%
\end{array}
$%
\caption{$\protect\eta $ versus $f(\protect\epsilon )$ for $T_{2}=100$, $%
T_{1}=80$, $P_{1}=10$, $P_{4}=3$, $k=-1$ (dashed line), $k=0$ (dotted line)
and $k=1$ (continuous line).}
\label{Fig10}
\end{figure}
\begin{figure}[tbp]
$%
\begin{array}{c}
\epsfxsize=7cm \epsffile{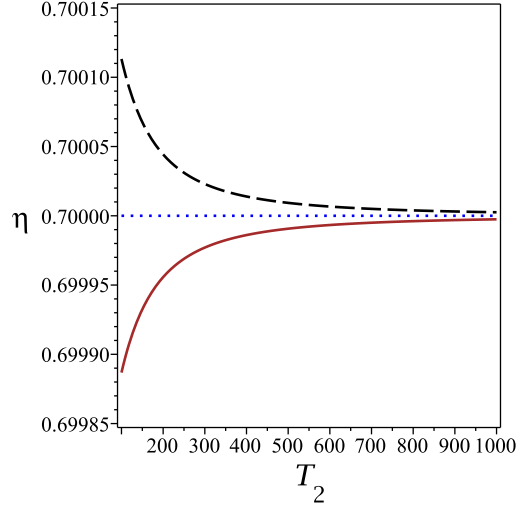}%
\end{array}
$%
\caption{$\protect\eta $ versus $T_{2}$ for $T_{1}=80$, $P_{1}=10$, $P_{4}=3$%
, $f(\protect\epsilon )=1.1$, $k=-1$ (dashed line), $k=0$ (dotted line) and $%
k=1$ (continuous line).}
\label{Fig11}
\end{figure}

Our results show that the rainbow function and the topological factor affect
efficiency. Considering that the horizon of black hole in gravity's rainbow
could be sphere, flat or hyperbolic, corresponding to $k=1$;$~0$ or $-1$,
respectively, we want to study the effects this factor on the engine's
efficiency. We found that the efficiency of black hole engines in this
gravity with hyperbolic horizon ($k=-1$) is higher than that of black holes
with flat horizon ($k=0$). Also, one finds that the spherical black holes ($%
k=1$) have the lowest efficiency. In other words, $\eta _{k=-1}>\eta
_{k=0}>\eta _{k=1}$. In order to have more details, we plot Figs. \ref{Fig9}%
, \ref{Fig10} and \ref{Fig11}.

The Fig. \ref{Fig9}, show that the efficiency ($\eta $) will monotonously
increase with the growth of pressure $P_{1}$. In other words $\eta $ will
approach to the maximum efficiency, i.e. the Carnot efficiency is allowed by
thermodynamics laws. In the high pressure limit ($P^{\ast }$), we have
\begin{equation}
\underset{P_{1}\rightarrow P^{\ast }}{\lim }\eta =1.
\end{equation}

Another interesting result is related to the effects of rainbow function.
According to this fact that the rainbow function ($f^{2}\left( \varepsilon
\right) $) located in the denominator of the second term of Eq. (\ref{eta}),
therefore by increasing the value of rainbow function (or by increasing the
effects of gravity's rainbow), the engine efficiency decreases. In the
absence of gravity's rainbow ($f^{2}\left( \varepsilon \right) =1$), the
effects of topological factor on engine efficiencies are clear, but by
considering large quantities of rainbow function more than one ($f\left(
\varepsilon \right) >1$), the behavior of engine efficiencies for
topological black holes are the same (see Fig. (\ref{Fig10})). Also, there
is the same behavior for high temperature (see Fig. (\ref{Fig11})). In other
words, for $T_{2}>>T_{1}$, the engine's efficiencies for topological black
holes are the same, but for $T_{2}$ near to $T_{1}$ ($T_{2}\simeq T_{1}$),
the differences are clear.

\section{Closing Remarks}

In this paper, we considered gravity's rainbow in order to include UV limit
in general relativity. First, we obtained solutions and showed that these
solutions may be interpreted as black holes with two horizons, extreme black
hole (one horizon) or naked singularity (without horizon). It is notable
that, the dependency of all constants on energy functions was considered.
The asymptotical behavior of obtained solutions was energy dependent (a)dS.
Then, we calculated thermodynamic and conserved quantities for these black
holes and showed that these quantities depend on rainbow functions, ($%
f(\varepsilon )$ and $g(\varepsilon )$), and satisfy the first law of
thermodynamics.

Next, we conducted a study regarding physical/nonphysical black holes
(positivity/negativity of temperature) and thermal stability of the
solutions. Our results showed that the physical limitation points depend on
topological factor ($k$), the modified cosmological constant ($\Lambda
(\varepsilon )$) and also the total charge ($Q$). We found that these black
holes could have three different behaviors;

(i) two roots for the temperature and the heat capacity, so that, the
obtained black holes were physical and enjoy thermal stability when their
entropy are in range $S_{1}<S<S_{{div}}$ (where $S_{1}$ and $S_{{div}}$ are
first root and divergency point).

(ii) one root or a critical entropy, $S_{1}$. Here, we come across two
different situations. In this case, we faced with physical and stable
topological charged black holes when entropy was bigger than critical
entropy ($S>S_{1}$).

(iii) no root. In this case, the obtain black holes were non-physical.

In order to study the phase transition critical points or divergence points,
we investigated denominator of the heat capacity. We found that there were
three different cases available for these black hole:

(i) two divergencies ($S_{1}^{\ast }$ and $S_{2}^{\ast }$). There was a
phase transition between the black holes with small entropy (or small black
holes) to large entropy (or large black holes) and vise versa. This phase
transition is related to first order transition.

(ii) one divergency ($S_{1}^{\ast }$). In this case, we encountered with two
different cases. \textit{CaseI}: existence of a phase transition from the
unstable black holes with large entropy to the stable black holes with small
entropy. \textit{CaseII}: no phase transition.

(iii) no divergency.

In addition, we used the geometrical thermodynamics for studying both the
physical limitation and phase transition critical points of topological
charged black holes in gravity's rainbow. It was shown that the divergencies
of the curvature scalar of the HPEM metric exactly coincide with both the
physical limitation and phase transition critical points of system. In other
words, the divergencies of heat capacity and its root are matched with the
divergencies of Ricci scalar. It is worthwhile to mention that the behavior
of Ricci scalar around its divergence points for roots and phase transition
critical points were different. It means that there were characteristic
behaviors that enable one to recognize the divergence point of Ricci scalar
related to the root of heat capacity or temperature (in this case the
divergencies around the physical limitation points were from positive
infinity to negative infinity and vise versa) from the divergence points of
Ricci scalar related to the divergencies of heat capacity (in this case the
divergencies around phase transition critical points were similar to $%
\Lambda $ which called $\Lambda $ divergencies). Therefore, one is able to
point out that a physical limitation and phase transition critical points
occurred in the divergencies of thermodynamical Ricci scalar of the HPEM
metric with different distinctive behaviors.

Finally, by considering the heat engine for black hole which was proposed by
Johnson, we studied the efficiency of heat engines for topological black
holes in gravity's rainbow. We found that the efficiency of heat engine
depends on topological factor and rainbow function ($f(\varepsilon )$). We
extracted some interesting results about the efficiency of heat engines for
these black holes as;

(i) the efficiency for black holes with hyperbolic horizon was bigger than
that of black holes with flat horizon ($k=0$), and the spherical black holes
($k=1$) had the lowest efficiency ($\eta _{k=-1}>\eta _{k=0}>\eta _{k=1}$).

(ii) with increasing pressure $P_{1}$, the efficiency reach to the maximum
efficiency, i.e. the Carnot efficiency is allowed by thermodynamics laws. In
fact, in the high pressure limit ($P^{\ast }$), $\underset{P_{1}\rightarrow
P^{\ast }}{\lim }\eta =1$.

(iii) in the absence of gravity's rainbow ($f\left( \varepsilon \right) =1$%
), the effects of topological factor on engine's efficiencies were
diffident, but for the large values of rainbow function ($f\left(
\varepsilon \right) >1$), the behavior of engine's efficiencies for
different topological black holes were the same.

(iv) for $T_{2}>>T_{1}$, the engine's efficiencies for different topological
black holes were the same, but for $T_{2}$ near to $T_{1}$ ($T_{2}\simeq
T_{1}$), the differences were clear.

\begin{acknowledgements}
This work has been supported financially by Research Institute for
Astronomy and Astrophysics of Maragha (RIAAM) under research
project No. 1/5440-22.
\end{acknowledgements}

\begin{center}
Appendix: phase transition and the heat engine efficiency
\end{center}

\begin{figure}[tbp]
$%
\begin{array}{cc}
\epsfxsize=6cm \epsffile{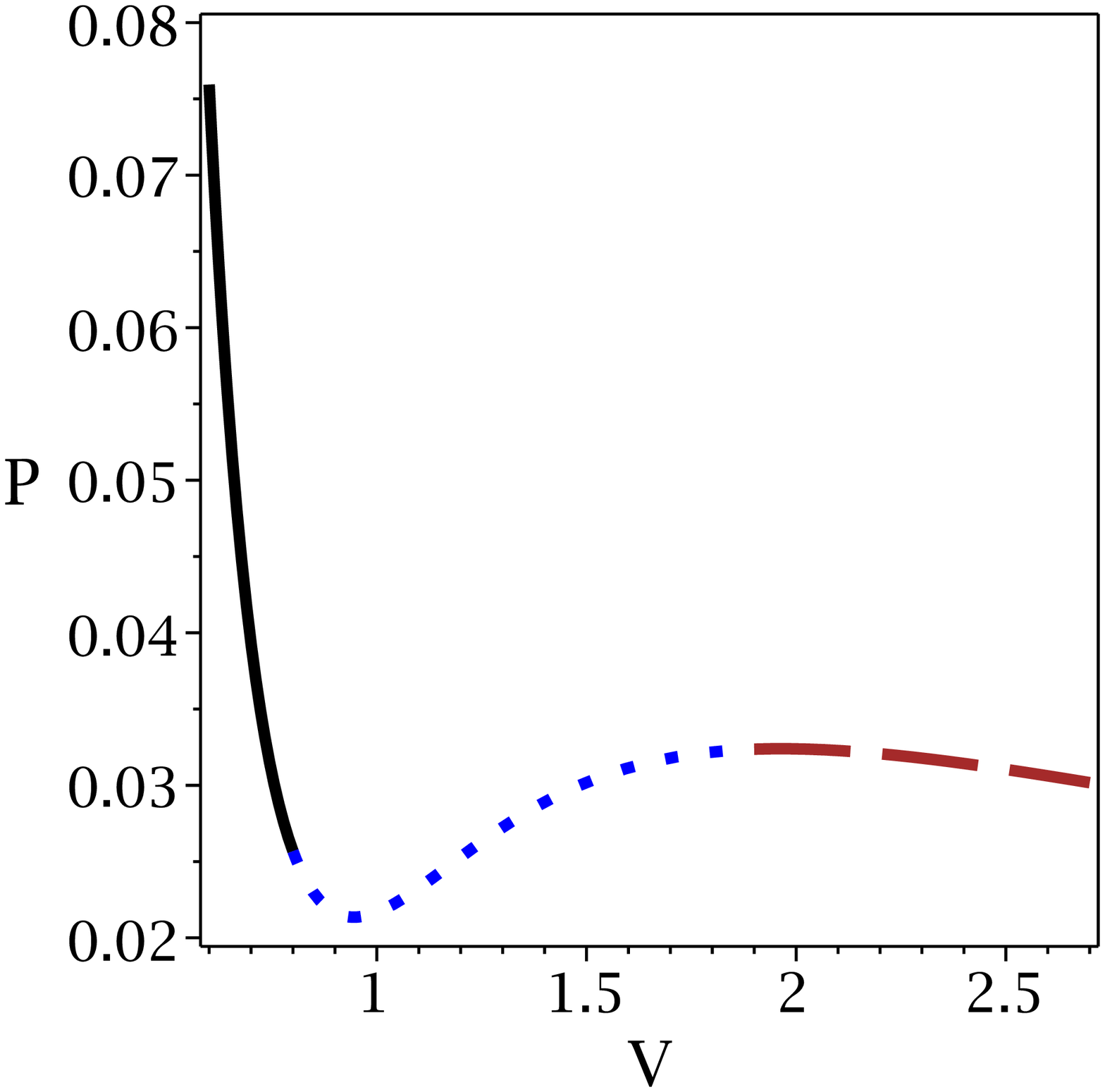} & \epsfxsize=6cm \epsffile{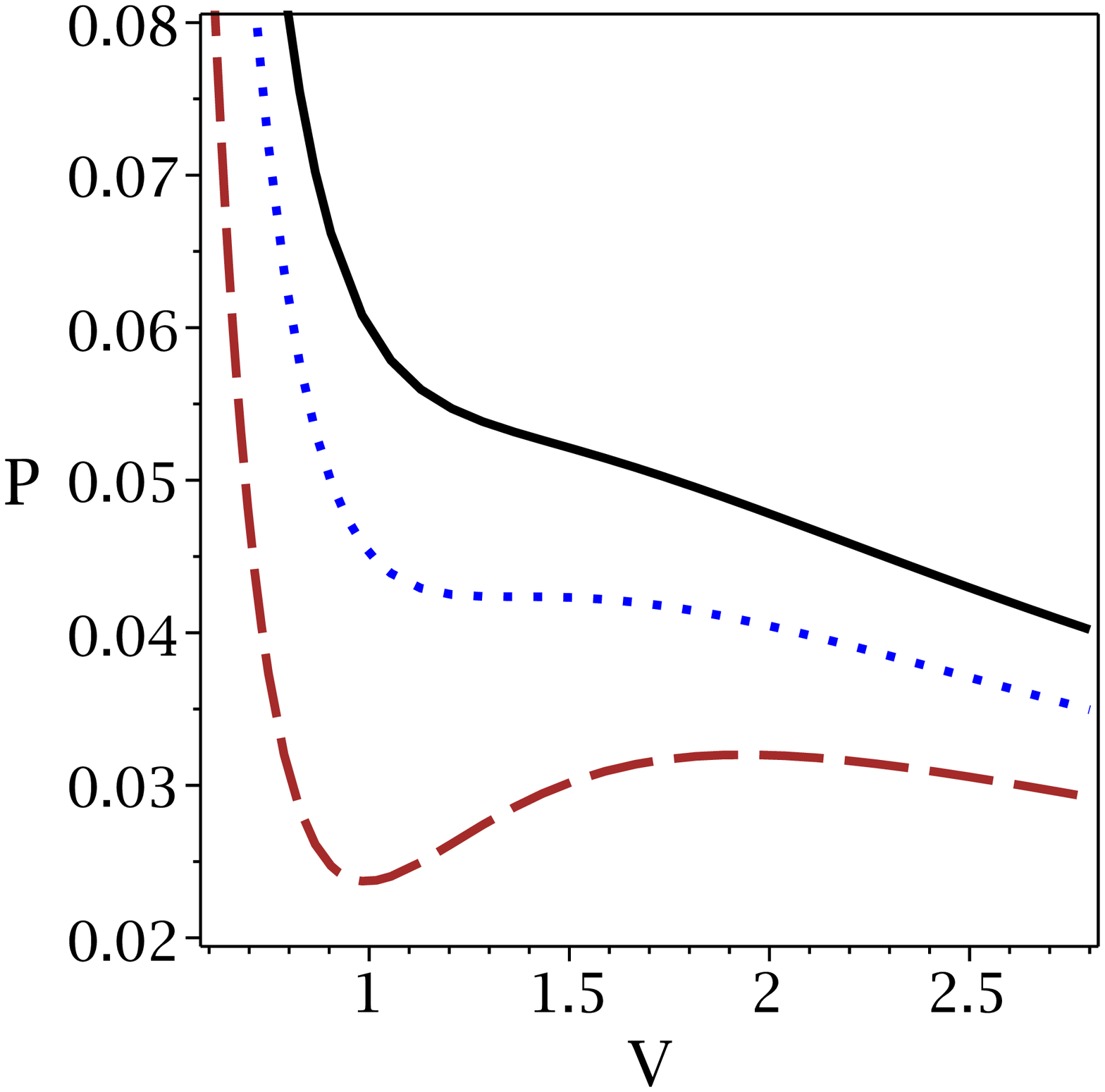}%
\end{array}
$%
\caption{\textbf{Left panel:} $P-V$ diagram for $T<T_{c}$. \newline
\textbf{Right panel:} $P-V$ diagrams for $T>T_{c}$ (continuous line), $%
T=T_{c}$ (doted line) and $T<T_{c}$(dashed line).}
\label{Fig12}
\end{figure}

Here, we are going to give more information about $P-V$\ diagram of a black
hole in different temperature. We have two cases for different temperature
in Fig. \ref{Fig12}.

CaseI: left panel of Fig. \ref{Fig12} shows that there are three areas for
black holes which encounter with a phase transition when $T\leq T_{c}$\ ($%
T_{c}$\ is critical temperature). The first area is related to high pressure
(continuous line in left panel in Fig. (\ref{Fig12})). The black hole in
this area has small radius and it is called small black hole (SBH) region.
The second area is related to unstable phase (dotted line in left panel in
Fig. (\ref{Fig12})). The third area is related to low pressure case (dashed
line in left panel of Fig. (\ref{Fig12})). In this area, black holes have
large radius, and it is known as large black hole (LBH) region.

Case II: for the temperatures more than critical temperature ($T>T_{C})$,
the phase transition and the second area are disappeared (see right panel of
Fig. (\ref{Fig12}) for more details). In phase transition point, the
transition takes place between SBH to LBH. The opposite could also take
place in the case black holes horizon shrinking (the LBH to the SBH).

In this work we considered the high temperature for investigating heat
engine of black hole in gravity's rainbow. Depending to the critical
temperature of our system, this temperature may include phase transition
area in the $P-V$\ plane or not.

In addition, the relation between the heat engine efficiency and $T-S$\
diagram have been evaluated in ref. \cite{WeiHeat}.


\end{document}